\documentclass{article}
 \pdfoutput=1 
\usepackage{spconf,amsmath,graphicx}
\usepackage{url}
\usepackage{color}

\usepackage{tikz,tikzscale}
\usepackage{pgfplots}
\pgfplotsset{compat=1.17} 
\usepackage{arydshln}
\usepackage{paralist}

\usepackage{textcomp}
\usepackage[absolute]{textpos}

\newcommand{\copyrightstatement}{
    \begin{textblock}{15}(0.5,0.3)    
         \noindent
         \centering
         \textblockcolour{white}
         \footnotesize
         \copyright 2022 IEEE. Personal use of this material is permitted. Permission from IEEE must be obtained for all other uses, in any current or future media, including reprinting/republishing this material for advertising or promotional purposes, creating new collective works, for resale or redistribution to servers or lists, or reuse of any copyrighted component of this work in other works
    \end{textblock}
}

\title{OPTIMIZED DECODING-ENERGY-AWARE ENCODING IN PRACTICAL VVC IMPLEMENTATIONS}
	\name{\setlength{\fboxsep}{0pt}\setlength{\fboxrule}{0pt}\framebox{
		   \begin{minipage}{\linewidth}\center
                      Matthias Kr\"anzler$^{\star }$ \ Adam Wieckowski$^{\dagger}$ \ Geetha Ramasubbu$^{\star}$ \ Benjamin Bross$^{\dagger}$ \newline Andr\'e Kaup$^{\star}$ \ Detlev Marpe $^{\dagger}$ \ Christian Herglotz$^{\star}$
                   \end{minipage}
                   }
        } 
  \address{$^{\star}$Multimedia Communications and Signal Processing, \\ Friedrich-Alexander-Universität Erlangen-Nürnberg (FAU), Erlangen, Germany \\
      $^{\dagger}$Video Communication and Applications Department, \\ Fraunhofer Heinrich-Hertz Institute (HHI), Berlin, Germany}

\begin{document}
\ninept
\copyrightstatement

\maketitle

\begin{abstract}

The optimization of the energy demand is crucial for modern video codecs. Previous studies show that the energy demand of VVC decoders can be improved by more than 50\% if specific coding tools are disabled in the encoder. However, those approaches increase the bit rate by over 20\% if the concept is applied to practical encoder implementations such as VVenC. Therefore, in this work, we investigate VVenC and study possibilities to reduce the additional bit rate, while still achieving low-energy decoding at reasonable encoding times. We show that encoding using our proposed coding tool profiles, the decoding energy efficiency is improved by over 25\% with a bit rate increase of less than 5\% with respect to standard encoding. Furthermore, we propose a second coding tool profile targeting maximum energy savings, which achieves 34\% of energy savings at bitrate increases below 15\%. 
\end{abstract}
\section{Introduction}
\label{sec:intro}

Over the past decade, we have witnessed the unprecedented success of online video services. A wide variety of applications such as on-demand streaming, social networks, and video conferencing tools has led to the fact that nowadays, more than three-quarters of the global Internet traffic constitute video data \cite{cisco20}. Consequently, research has put substantial efforts in developing new compression methods to reduce transmission bitrates, which led to highly sophisticated video codecs such as High-Efficiency Video Coding (HEVC) \cite{Sullivan12} and Versatile Video Coding (VVC) \cite{Bross21}. 

At the same time, the global energy demand for such online video services increased rapidly. As a consequence, in the year 2018, it was estimated that $1\%$ of the global greenhouse gas emissions were caused by online video \cite{ShiftFull19}. Hence, research targeting energy efficiency in online video is crucial for the sustainable use of this technology in the future. 

To increase the energy efficiency of online video applications, researchers developed a multitude of solutions. For example, an in-depth analysis of the power consumption of smartphones during video playback was performed \cite{Carroll13}, which were used to develop power models \cite{Herglotz20} that help reduce the power consumption when changing sequence parameters such as the resolution \cite{Herglotz19b}. 
Furthermore, within the GREEN-MPEG standard, multiple signaling methods were standardized allowing transmitters and receivers to exchange messages and requests that enable to decrease the overall power consumption \cite{GREEN-MPEG}. 

One branch of this research targets the optimization of the decoder energy consumption, where several methods optimizing the decoding complexity were proposed for HEVC~\cite{Mallikarachchi20,Correa18,Herglotz19}. Recently, with the standardization of the next-generation video codec VVC, approaches to reduce the energy consumption of the VVC decoder were developed and analyzed in detail~\cite{Kraenzler20,Kraenzler21}. These methods exploit the encoder configuration to reduce the decoder's energy consumption. Unfortunately, these methods were developed and optimized for the reference implementation of the VVC codec \cite{VTM}, which is not used in practical applications of online video services. 

In this paper, we refine this encoder configuration for a practical implementation of VVC targeting real-world scenarios. We choose VVenC \cite{Wieckowski21,VVenC} as an encoder and VVdeC as a decoder \cite{Wieckowski20,VVdeC}. We perform an in-depth analysis of the impact of VVenC configurations on the output bitrate and the decoding energy and propose multiple coding tool profiles (CTP) leading to decoding energy savings with respect to standard encoding. The novel contributions of this paper are as follows: 
\begin{compactitem}
\item Analysis of the energy consumption of a real-world VVC decoder, 
\item Development of a new encoding configuration targeting energy-efficient VVC decoding with the constraint of a minimum encoding time. 
\end{compactitem}

This paper is organized as follows. In Section~\ref{sec:setup}, we introduce the encoder as well as the decoder implementation used for this study and explain the energy measurement setup. Section~\ref{sec:ConfigDescription} presents the derivation of the optimized encoder configuration for decoding-energy optimized encoding. Then, Section~\ref{sec:eval} discusses the results from the evaluation and the validation of the proposed encoder configuration. Finally, Section~\ref{sec:concl} concludes this paper.

\section{Setup}
\label{sec:setup}

\subsection{Encoder and Decoder Implementation} 

Most literature concerning VVC is based on the VVC Test Model (VTM) implementation. Both the encoder and the decoder provided by VTM were developed on a common software basis to be used in the standardization process for codec exploration experiments. For this reason, they lack practical options as well as thorough optimization. Being based on a single software basis, especially the decoder cannot be properly optimized, since the data structures and algorithms also have to support the more general encoding scenarios.

The researchers at Fraunhofer HHI have developed optimized software implementations of both VVC encoder (VVenC) \cite{Wieckowski21} and decoder (VVdeC)~\cite{Wieckowski20} for practical uses. They were released as open source software towards the end of 2020. Both projects are based on the VTM software but are heavily optimized.

\subsubsection{VVdeC software decoder}

The Fraunhofer Versatile Video Decoder (VVdeC)~\cite{Wieckowski20} is an optimized software decoder library that provides low-level optimizations and multi-threading. It supports the full scope of the Main10 VVC profile. The software can be cross-compiled to provide VVC playback support on a variety of architectures, including x86 and ARM architectures (both 32- and 64-bit) as well as to a WebAssembly for browser support. VVdeC has been used to provide VVC playback capabilities in several demonstrators and can be easily integrated into multi-media frameworks \cite{VVCandOSS}. Version 1.3.0 of the decoder has been released in December 2021, on which the following measurements are based.

VVdeC does not contain logic to properly time picture output, but rather tries to decode the incoming packages as fast as possible, utilizing as much of its assigned resources as possible. It is up to the implementing library to ensure proper timing of picture display after decoding.

\subsubsection{VVenC software encoder}

The Fraunhofer Versatile Video Encoder (VVenC) \cite{Wieckowski21} is an optimized software encoder library as well as a standalone encoder application. We use the latest version 1.3.1 from December 2021 for our studies.

VVenC provides a whole range of practical features, including five predefined presets: \textit{faster}, \textit{fast}, \textit{medium}, \textit{slow}, and \textit{slower}. The \textit{slower} preset provides all of VTM's efficiency and the remaining four presets trade off BDR losses for runtime reduction.

The encoder further allows multi-threaded operation with runtime scaling. It includes single- and two-pass rate control modes allowing encoding at a specific rate rather than target quality.
VVenC includes a subjective quality optimization based on the XPNSR measure \cite{Helmrich20}, which has been proven to be very effective in the JVET verification tests, allowing VVenC in \textit{medium} preset to reach VTM's subjective quality \cite{JVET_T2020,JVET_U0119}.
The encoder further has application-specific tools allowing its use with screen-content coding, HDR content, streaming applications, and more.

\subsection{Energy Measurement Setup} 
The energy demand of the decoding process is determined with the help of two consecutive measurements, as explained in \cite{Herglotz18}. Initially, the energy consumed during the decoding process is measured. Then, the energy consumed in idle mode over the same duration as the decoding process is measured. In the end, the decoding energy is the difference between these two measurements.

We performed the energy measurements on a desktop PC with an Intel i7-8700 CPU using its integrated power meter, Running Average Power Limit (RAPL). Due to the influence of noise and background processes, we perform a series of measurements and validate the results with a confidence interval test for statistical correctness as described in~\cite{Kraenzler21}. As parameters for the confidence interval, we chose $\alpha=0.99$ and $\beta=0.01$. By doing so, we ensure that the measured decoding energy has a maximum energy deviation of 1\% from the actual energy consumed. 

For bitrate efficiency evaluation of a coding tool profile (CTP), which indicates the usage of each coding tool during encoding~\cite{Kraenzler21}, we use the commonly used Bj{\o}ntegaard Delta (BD) metric. In this work, we refer Bj{\o}ntegaard Delta Bitrate (BDR) to the bit rate increase/savings for the same $\textrm{PSNR}_{\textrm{YUV}}$ level. For the calculation of BDR and PSNR, we use the guidelines described in~\cite{Strom21}. Furthermore, we use the Bj{\o}ntegaard Delta Decoding Energy (BDDE) to evaluate the impact of the decoding energy demand for the same $\textrm{PSNR}_{\textrm{YUV}}$ level, correspondingly. For the calculation of BDDE, we substitute the bitrate with the decoding energy demand as was done in~\cite{Kraenzler21}. For each BD metric in this work, the reference is the \textit{medium} preset of VVenC.

\begin{table}
\caption{Evaluated configurations on top of VVenC \textit{medium}. \textit{EE} represents the reduced CTP according to \cite{Kraenzler21}. The CTP configurations \textit{v2}-\textit{v8} try to compensate the bitrate loss of \textit{EE} with minimal decoding energy increase.}
\begin{center}
\renewcommand{\arraystretch}{1.3}
\begin{tabular}{c | l}
CTP & Options \\
\hline
 EE (\textit{v1}) & \multicolumn{1}{p{6.5cm}}{\texttt{--Affine=0 --MIP=0 --ISP=0 --LMChroma=0 --BDOF=0 --DMVR=0 --PROF=0 --SbTMVP=0 --SMVD=0 --LFNST=0 --LoopFilterDisable --EDO=0 --SAO=0 --LMCS=0 --BCW=2 --CCALF=0 --ALF=0}} \\
 v2     & EE \& \texttt{--ALF --CCLAF --ALFSpeed} \\
 v3     & EE \& \texttt{--BDOF --DMVR} \\
 v4     & EE \& \texttt{--MMVD=1 --Geo=1 --AMVR=1} \\
 v5     & \multicolumn{1}{p{6.5cm}}{EE \& \texttt{--MaxMTTDepthISliceL=3 \newline --MaxMTTDEepthISliceC=3 \newline --MaxMTTDepth=2}} \\
 v6     & EE \& \texttt{--Affine=2 --PROF} \\
 v7     & EE \& \texttt{--LoopFilterDisable=0 --EDO=2} \\
 v8     & EE \& \texttt{--LMChroma} \\
 v58    & v5 \& v8 \\
 v258   & v2 \& v5 \& v8 \\  
 v2568  & v2 \& v5 \& v6 \& v8 \\
\end{tabular}
\renewcommand{\arraystretch}{1.0}

\end{center}
\label{tab:opts}
\end{table}

\section{Decoding-Energy Optimized Coding Tool Profiles}
\label{sec:ConfigDescription}
For this paper, we have chosen the \textit{medium} preset as a baseline since it includes most of VVC's tools but also runs at a fraction of VTM's runtime. VVenC's subjective optimization has been enabled~\mbox{(\texttt{--PerceptQPA=1})} in the experiments to simulate more realistic bitstream characteristics with varying quantization parameter values across frames and blocks. For our experiments, we first reduce the \textit{medium} preset to the settings from the \textit{Energy Efficient (EE)} CTP, which was determined for VTM using a Design Space Exploration in \cite{Kraenzler21}. Note that the~\textit{EE} CTP in this paper is reduced to the scope of the~\textit{medium} preset, which means that the partitioning is reduced and some coding tools have a reduced search space. In~\cite{Kraenzler21}, the \textit{EE} profile showed significant energy savings of up to 50\% for two decoder implementations. However, the authors also observed a significant BDR increase.

Based on the CTP \textit{EE}, we selectively enable encoder options and measure the individual impact on the BDR, the decoding energy, and the encoding time. Table \ref{tab:opts} shows all tested options and different combinations based on the most efficient candidates. The reasoning for the selected options is as follows.  
Next to obtaining a minimum decoding energy, tools are added in such a way that simultaneously, the bitrate and the encoding times are as close as possible to the bitrate and the encoding times of the \textit{medium} preset, respectively.

The candidate tested extensions of \textit{EE}, \textit{v2}--\textit{v8}, which were not available in the VTM encoder~\cite{Kraenzler21}, are shown in Table~\ref{tab:opts}. The reasoning behind the chosen candidates is threefold. Firstly, \textit{v2} is a configuration option of VVenC designed to reduce encoding runtime of the adaptive loop filter (ALF) by only applying it on referenced frames (i.e. every second frame in the temporal structure used by the encoder). Such a configuration already provides most of the gains of ALF and cross component ALF (CCALF). Because of the application to every second frame only, it is assumed that the decoding energy requirement of this decoder-heavy filter is also halved. Secondly, since the tests are performed on a practical optimized VVC decoder implementation, VVdeC, \textit{v3} and \textit{v6}--\textit{v8} are used to validate if the decoding energy behavior measured in VTM is still negative for an optimized implementation. Finally, because of the way the presets are derived in VVenC, some of the coding tools are enabled in a reduced scope. For \textit{v4}, we extend the search space of the motion derivation tools merge with motion vector difference (MMVD) and adaptive motion vector refinement (AMVR). Furthermore, an alternative bi-prediction blending method for the geometric partitioning mode (Geo) is used. For \textit{v5}, we apply a deeper partitioning on the encoder, which increases the multi type tree depth of the encoder. 

Lastly, a detailed description of the used coding tools can be found in~\cite{Bross21}. For the CTPs \textit{v58}, \textit{v258}, and \textit{v2568}, we combine the corresponding modifications as shown in the second column of Table~\ref{tab:opts}.

\begin{table} 
\caption{CTP results for HHI HD SDR test set~\cite{JVET_Q0791} relative to the VVenC \textit{medium} preset.}
\begin{center}
\renewcommand{\arraystretch}{1.2}
\begin{tabular}{c | r : r : r}
CTP        & Enc. Time & ~~~BDR~~~ & ~~BDDE~~ \\
\hline
 \textit{EE (v1)} & 65.01\%      &  10.42\%  & -31.99\% \\
 \textit{v2}     & 65.74\%      &   6.01\%  & -24.16\% \\
 \textit{v3}     & 70.14\%      &   9.28\%  & -20.33\% \\
 \textit{v4}     & 107.44\%     &   9.39\%  & -32.88\% \\
 \textit{v5}     & 112.81\%     &   8.25\%  & -31.95\% \\
 \textit{v6}     & 67.63\%      &  10.20\%  & -31.58\% \\
 \textit{v7}     & 66.41\%      &   8.46\%  & -27.70\% \\
 \textit{v8}     & 65.79\%      &   8.15\%  & -31.57\% \\
 \textit{v58}    & 115.16\%     &   5.95\%  & -30.29\% \\
 \textit{v258}   & 115.09\%     &   2.11\%  & -23.24\% \\
 \textit{v2568}  & 119.28\%     &   1.86\%  & -21.67\% \\
\end{tabular}
\renewcommand{\arraystretch}{1.0}
\end{center}
\label{tab:hhires}
\end{table}

All CTPs have been first tested on the HHI HD SDR test set \cite{JVET_Q0791} in JVET random access configuration \cite{JVET_N1010}. By doing so, we can validate our results on another set, which is the JVET common test condition set~\cite{JVET_N1010}. The results in comparison to VVenC \textit{medium} are presented in Table~\ref{tab:hhires}. The first reduced configuration \textit{EE} provides high decoding energy reduction but also shows a high bitrate loss of 10.42\% on the HHI test set. Some of the tested options reduce the bitrate loss substantially without increasing the decoding energy (\textit{v5}, \textit{v8}). The reduced ALF configuration \textit{v2} provides a very high BDR gain, while slightly increasing the decoding energy. Configurations \textit{v3} and \textit{v7} confirm the findings in \cite{Kraenzler21} that tools even with VVdeC's optimized implementation should not be switched on/off. For \textit{v4}, we find that both BDDE and BDR are improved. However, the encoding time doubles compared to \textit{EE}. Since the configuration \textit{v5} has a higher improvement in terms of BDR and a similar increase of the encoding time, we select configuration \textit{v5} over \textit{v4} for the combination with other configurations. By doing so, we limit the encoding time to a reasonable level.

\definecolor{C1}{HTML}{DBAC0B}
\definecolor{C2}{HTML}{009124}
\definecolor{C3}{HTML}{1129D9}
\definecolor{C4}{HTML}{00BD47}
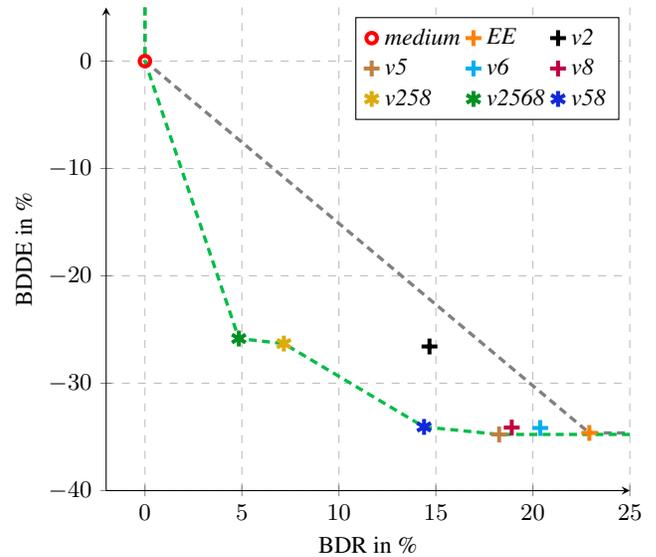
\begin{figure}[!t]
\begin{center}
\begin{tikzpicture}
\begin{axis}[
	width=0.48\textwidth,
	height =0.45\textwidth,
    xlabel={BDR in \%},
    ylabel={BDDE in \%},
    legend cell align={left},
    xmin=-2, xmax=25,
    ymin=-40, ymax=5,
    axis lines = left,
    ymajorgrids=true,
    xmajorgrids=true,
    grid style=dashed,
    legend columns=3,]
    
	\addplot[only marks,color=red,mark=o,line width=1.5pt,mark size = 2pt,]
    coordinates {(0,0)};
    \addlegendentry{\textit{medium}}

	\addplot[only marks,color=orange,mark=+,line width=1.5pt,mark size = 3pt,]
    coordinates {(22.91,-34.63)};
    \addlegendentry{\textit{EE}}

	\addplot[only marks,color=black,mark=+,line width=1.5pt,mark size = 3pt,]
    coordinates {(14.67,-26.58)};
    \addlegendentry{\textit{v2}}
  
	\addplot[only marks,color=brown,mark=+,line width=1.5pt,mark size = 3pt,]
    coordinates {(18.26,-34.77)};
    \addlegendentry{\textit{v5}}
    
	\addplot[only marks,color=cyan,mark=+,line width=1.5pt,mark size = 3pt]
    coordinates {(20.38,-34.17)};
    \addlegendentry{\textit{v6}}    
    
    \addplot[only marks,color=purple,mark=+,line width=1.5pt,mark size = 3pt,]
    coordinates {(18.91,-34.13)};
    \addlegendentry{\textit{v8}}
    
   \addplot[only marks,color=C1,mark=asterisk,line width=1.5pt,mark size = 3pt]
    coordinates {(7.16,-26.31)};
    \addlegendentry{\textit{v258}}
    
   \addplot[only marks,color=C2,mark=asterisk,line width=1.5pt,mark size = 3pt]
    coordinates {(4.84,-25.84)};
    \addlegendentry{\textit{v2568}}

    \addplot[only marks,color=C3,mark=asterisk,line width=1.5pt,mark size = 3pt]
    coordinates {(14.39,-34.07)};
    \addlegendentry{\textit{v58}}

\draw [ 
   color=gray,mark=none,densely dashed, mark size=2pt,line width=1.25pt] 
   plot coordinates {(0,5) (0,0) (22.91,-34.63) (25.0,-34.63)  };

\draw [ 
   color = C4,densely dashed,mark=none, mark size=2pt,line width=1.25pt] 
   plot coordinates {(0,5) (0,0) (4.84,-25.84) (7.16,-26.31) 
  (14.39,-34.07) (18.26,-34.77) (25.0,-34.77)  };
    
\end{axis}
\end{tikzpicture}
\end{center}
\caption{Results of the proposed CTPs for the JVET CTC set with the corresponding BDR (vertical axis) and BDDE (horizontal axis) values relative to the \textit{medium} preset. The gray curve corresponds to the Pareto front of the literature, and the green curve to our proposed Pareto front.}
\label{fig:ValidationPlot}
\end{figure}

\section{Validation}
\label{sec:eval}

\setlength{\tabcolsep}{3.75pt}
\begin{table*}[!ht]
\caption{Evaluation of the configuration \textit{EE} and the proposed CTPs for the JVET CTC set. For each CTP, the BDR and BDDE value is calculated in relation to the \textit{medium} preset. Each BD value is given in \%.}
\vspace{0.25cm}
 \label{tab:EvaluationJVET}
\begin{small}
\renewcommand{\arraystretch}{1.2}
\begin{tabular}{l || c : c || c : c | c : c | c : c | c : c || c : c | c : c | c : c  }
	& \multicolumn{2}{c||}{\textit{EE (v1)}} & \multicolumn{2}{c|}{\textit{v2}} & \multicolumn{2}{c|}{\textit{v5}} & \multicolumn{2}{c|}{\textit{v6}} & \multicolumn{2}{c||}{\textit{v8}} & \multicolumn{2}{c|}{\textit{v58}} & \multicolumn{2}{c|}{\textit{v258}} & \multicolumn{2}{c}{\textit{v2568}} \\     
	&	 BDR & BDDE &	 BDR &  BDDE &	 BDR &  BDDE &	 BDR &  BDDE &	 BDR &  BDDE 	& BDR &  BDDE 	& BDR &  BDDE &	 BDR &  BDDE \\
\hline
A1	& 	32.24 & -45.19 & 	21.78 & -32.11 & 	27.83 & -44.92 & 	30.26 & -44.40 & 	20.73 & -44.77 & 	16.55 & -44.29 & 	8.66 & -32.53 & 	6.75 & -31.73 \\
A2	& 	32.97 & -47.90 & 	20.23 & -35.25 & 	27.05 & -48.05 & 	25.76 & -46.68 & 	30.11 & -47.64 & 	24.35 & -47.65 & 	12.87 & -36.07 & 	6.45 & -34.76 \\
B	& 	26.37 & -35.40 & 	15.88 & -27.06 & 	21.91 & -35.51 & 	22.26 & -34.89 & 	23.35 & -35.04 & 	18.98 & -34.86 & 	9.69 & -26.92 & 	6.20 & -26.32 \\ 	
C	& 	16.69 & -28.56 & 	11.19 & -22.20 & 	12.49 & -28.76 & 	16.25 & -28.46 & 	14.03 & -27.62 & 	9.93 & -27.81 & 	4.96 & -21.14 & 	4.40 & -21.03 \\
D	& 	23.93 & -50.37 & 	14.14 & -44.39 & 	19.30 & -50.42 & 	22.84 & -50.11 & 	21.34 & -49.20 & 	16.72 & -49.47 & 	7.52 & -43.08 & 	6.22 & -42.85 \\ 	
F	& 	15.24 & -18.70 & 	11.21 & -13.03 & 	9.48 & -19.24 & 	13.95 & -18.50 & 	11.41 & -18.63 & 	5.84 & -18.43 & 	2.31 & -12.71 & 	1.08 & -12.65 \\ 	
\hline
Avg	& 	22.91 & -34.63 & 	14.67 & -26.58 & 	18.26 & -34.77 & 	20.38 & -34.17 & 	18.91 & -34.13 & 	14.39 & -34.07 & 	7.16 & -26.31 & 	4.84 & -25.84 \\ 	
\end{tabular}
\renewcommand{\arraystretch}{1.0}
 \end{small}
\end{table*}

In the following, we will validate a subset of CTPs from Section~\ref{sec:ConfigDescription} using the JVET CTC set. Thereby, we evaluate whether the improvements are also present for another video set. We reduce the set of considered CTPs to \textit{v2}, \textit{v5}, \textit{v6}, \textit{v8}, \textit{v58}, \textit{v258}, and \textit{v2568} based on the results in Table~\ref{tab:hhires}.
CTP \textit{v6} has been included in the validation set, because it reduces the additional bitrate of the CTP \textit{v258} below 2\% for the CTP \textit{v2568}.  
In Table~\ref{tab:EvaluationJVET}, we provide the corresponding BDR and BBDE values for each class of the JVET CTC set and the average values over all sequences for each considered CTP. In Fig.~\ref{fig:ValidationPlot}, the validation of the proposed CTPs, the \textit{EE} CTP, and the reference \textit{medium} preset is shown. For the \textit{EE} CTP, we measure a BDDE of -34.63\% and a BDR of 22.91\%, which shall be both improved by our proposed CTPs. It is observed that the CTP \textit{v6} and {v8} provide a similar energy efficiency as \textit{EE} with a BDDE value of approximately -34\%. Furthermore, each CTP improves the BDR value of \textit{EE} by at least 2\%. For the CTP \textit{v5}, we determine that the BDDE is -34.77\% and lower than the corresponding value of the \textit{EE} CTP. Therefore, we can conclude that the partitioning depth of the encoder can improve the energy efficiency of VVC decoding, which was not studied before. The combination of the CTP \textit{v5} and \textit{v8} corresponds to the CTP \textit{v58}, which has a BDDE value of \mbox{-34.07\%} and BDR value of \mbox{14.39\%}. 

For the \textit{v2} CTP, we determine that with a BDR of \mbox{14.67\%}, the same improvement of the bit rate efficiency is achieved as for \textit{v58} (\mbox{14.39\%}), but the energy efficiency is lower (\textit{v2}: \mbox{-26.58\%}, \textit{v58}: \mbox{-34.07\%}). To improve the bit rate efficiency of \textit{v58} even more, we add the CTPs \textit{v6} and \textit{v8}. Consequently, the combined CTPs \textit{v258} and \textit{v2568} both optimize the bit rate efficiency significantly to \mbox{7.16\%} (\textit{v258}) and \mbox{4.84\%} (\textit{v2568}). Furthermore, the BDDE value is increased by less than \mbox{10\%} in relation to \textit{v8}, which shows that we can reduce the additional bit rate from \mbox{18.91\%} (\textit{v8}) to \mbox{4.84\%} (\textit{v2568}). In terms of the encoding time, we derived similar results as shown in Table~\ref{tab:hhires}. In Table~\ref{tab:EvaluationJVET}, we observe that the sequences of Class F have a significantly lower improvement of the energy demand in comparison to the other classes. This can be explained by the fact that those sequences contain screen content and thus, have different properties than natural sequences.

In Fig.~\ref{fig:ValidationPlot}, we show the improved Pareto Front with our proposed CTPs by a green line. It can be determined that the curve is shifted to the bottom left in relation to the gray curve, which corresponds to the Pareto front from literature. The only CTP that has a higher energy efficiency than \textit{EE} is \textit{v5}, which uses deeper partitioning in the encoder that is not available with VTM. As a result, if a low increase in bit rate is desired, we propose to use CTP \textit{v2568} for the improvement of the energy efficiency. For a high improvement of the energy efficiency, we propose to use the CTP \textit{v58}, which has an improvement of the BDR value of over \mbox{10\%} compared to CTP \textit{EE}.

To verify the results of the evaluation on another decoder implementation, we use the VTM decoder version 15.2~\cite{VTM}. We measured that the \textit{EE} CTP improves the energy of the \textit{medium} preset by \mbox{-46.68\%}. Again, the CTP \textit{v5} has the lowest energy demand with a BDDE of \mbox{-47.48\%}. For the CTPs \textit{v2}, \textit{v6}, and \textit{v8}, we observe that the energy efficiency is not improved compared to the \textit{EE} CTP. For \textit{v58}, we measure a BDDE of \mbox{-47.63\%}, and for \textit{v2568} of \mbox{-40.61\%}. Therefore, we conclude that the observation of VVdeC are also valid for the VTM decoder.

\section{Conclusion}
\label{sec:concl}

In this paper, we developed and optimized VVC encoder configurations targeting a low decoding energy at acceptable encoding times. The following changes lead to improvement of the bit rate efficiency: the enabling of a linear ALF, which is used on every second frame, a deeper partitioning, the usage of affine motion, and the usage of the coding tool CCLM. Furthermore, the combination of those modifications leads to more improvements. 

In summary, we propose two profiles to improve the results from literature in terms of compression efficiency. If a similar energy efficiency should be achieved, we propose to use the profile \textit{v58}, which reduces the energy demand by \mbox{-34.07\%} and increases the bitrate by \mbox{14.39\%}. If the compression efficiency shall be improved significantly, we propose to use the profile \textit{v2568}, which increases the bit rate by less than \mbox{5\%}, but reduces the energy demand by more than \mbox{25\%}. Based on the presented results, VVenC will include a decoding energy-optimized configuration along with other presets in the future.

In future work, we want to improve the energy efficiency for screen content video sequences. Furthermore, the influence of real-time decoding can be studied.

\bibliographystyle{IEEEbib}

\end{document}